\documentclass[aps,prl,twocolumn,superscriptaddress]{revtex4}
\usepackage{graphicx}
\usepackage{amsmath}				
\usepackage{amssymb}				
\usepackage{color}
\usepackage{pdfcolmk}
\newcommand{\bfm}{\mathbf}

\begin{document}

\title{Manipulation of the magnetic ground state of individual manganese phtalocyanines by coordination to CO molecules}

\author{A. Str\'{o}\.{z}ecka}
\affiliation{Institut f\"{u}r Experimentalphysik, Freie Universit\"at
Berlin, 14195 Berlin, Germany}
\author{M. Soriano}
\affiliation{Departamento de F\'{i}sica de la Materia Condensada,
Universidad Aut\'{o}noma de Madrid, Campus de Cantoblanco, 28049 Madrid,
Spain} \affiliation{Departamento de F\'{i}sica Aplicada, Universidad de
Alicante, 03690 San Vicente del Raspeig, Alicante,  Spain}
\author{J. I. Pascual}
\affiliation{Institut f\"{u}r Experimentalphysik, Freie Universit\"at
Berlin, 14195 Berlin, Germany}
\author{J. J. Palacios}
\affiliation{Departamento de F\'{i}sica de la Materia Condensada,
Universidad Aut\'{o}noma de Madrid, Campus de Cantoblanco, 28049 Madrid,
Spain}
\date{\today}

\begin{abstract}
We show that the magnetic state of individual manganese phthalocyanine
(MnPc) molecules on a Bi(110) surface is  modified when the Mn${2+}$
center coordinates to CO molecules adsorbed on top.  Using scanning
tunneling spectroscopy we identified this change in magnetic properties
from the broadening of a Kondo-related  zero-bias anomaly when the CO-MnPc
complex is formed. The original magnetic state can be recovered by
selective desorption of individual CO molecules. First principle
calculations show that the CO molecule reduces the spin of the adsorbed
MnPc from $S=1$ to $S=1/2$ and strongly modifies the respective screening
channels, driving a transition from an underscreened Kondo state to a
state of mix-valence.
\end{abstract}




\maketitle

Control over the magnetic moment of a molecule and its interaction with a
substrate is a key issue in the emerging field of molecular spintronics
\cite{Bogani:2008}. In metal-phthalocyanines and metal-porphyrins the
metal center is usually coordinatively unsaturated and presents a local
reactive site, which opens a unique possibility of controlling the
magnetic moment in-situ by external chemical stimuli
\cite{Gottfried:2009,Flechtner:2007,Hieringer:2011,Miguel:2011,Wackerlin:2010,Isvoranu:2011_Ammonia,Isvoranu:2011_CO}.
The axial coordination of small molecules like CO, NO or O$_2$ to those
complexes substantially alters their electronic properties, which has led
to a successful implementation of phthalocyanines  and porphyrins in gas
sensors \cite{Guillaud:1998,Bohrer:2007}. Studies specifically addressing
the magnetic state are however scarce. Only recently it has been shown
that the attachment of NO molecule to a cobalt-tetraphenylporphyrin
(CoTPP)  quenches its spin due to the oxidation process
\cite{Wackerlin:2010}.
 The general picture is however more complicated, as the chemical bond to the reactant molecule causes the redistribution of charge in the $d$-orbitals of the metal center and modifies the ligand field of the metal ion. This has critical consequences for the magnetic ground state of the complex \cite{Isvoranu:2011_Ammonia,Isvoranu:2011_CO}. On metal surfaces, the formation of a new ligand bond may additionally alter the hybridization of molecular and substrate states, thus affecting the electronic and magnetic coupling of the metal ion to the substrate \cite{Flechtner:2007,Hieringer:2011,Miguel:2011}. Understanding the response of these effects to the change in chemical coordination is crucial to gain the full control over the functionality of the magnetic system.

\begin{figure}
\includegraphics[width=0.90\columnwidth]{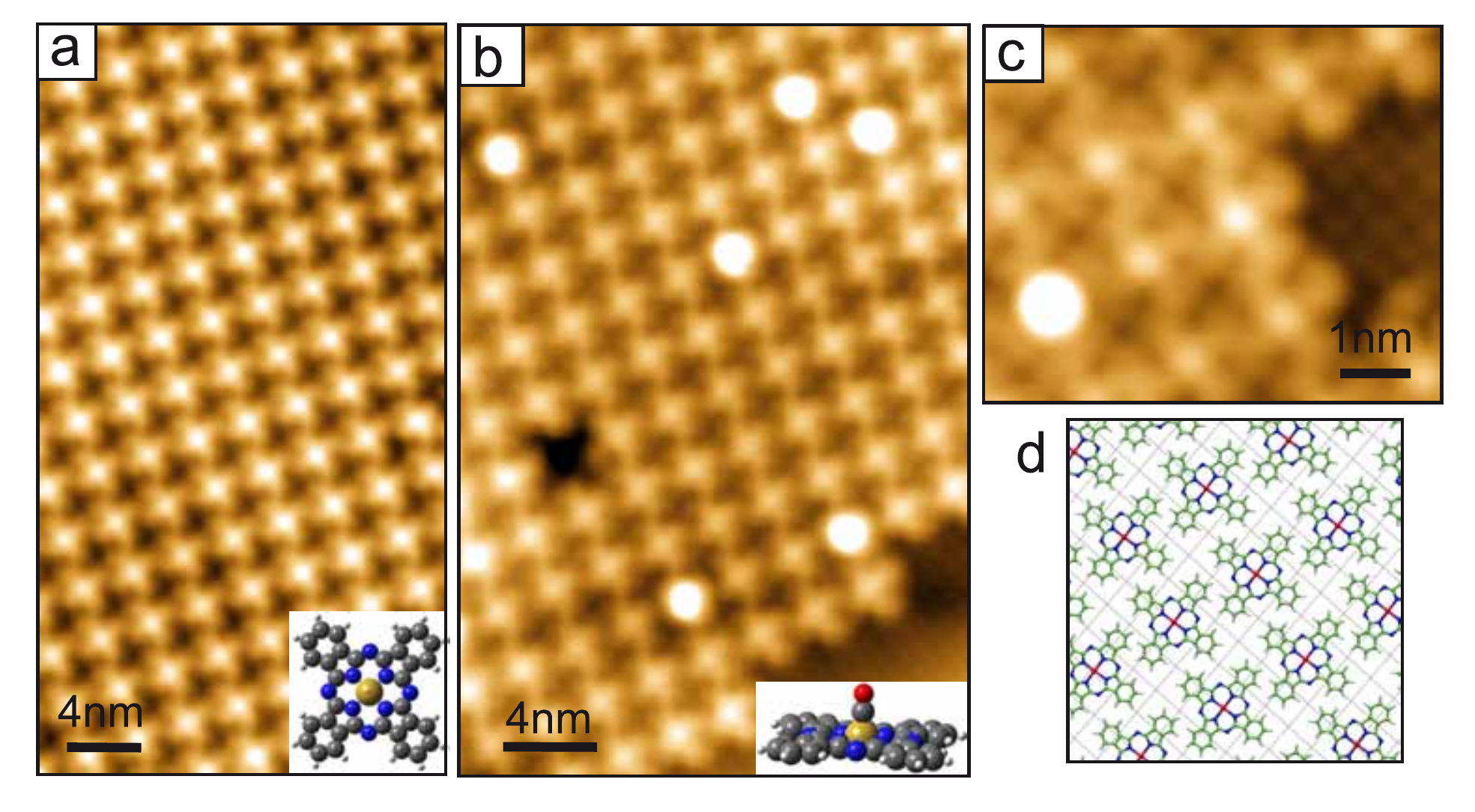}
\caption {Adsorption of MnPc and CO-MnPc on Bi(110). (a) STM image of highly ordered MnPc island ($I$=0.2\,nA, $V$=250\,mV, \cite{WSXM}).
(b) MnPc island after exposure to CO ($I$=0.1\,nA, $V$=180\,mV). CO-coordinated molecules can be distinguished by different apparent height. Inset images in (a) and (b) show schematic pictures of the chemical structure of MnPc and CO-MnPc. (c) High resolution STM image of bare and CO-ligated MnPc ($I$=0.1\,nA, $V$=-250\,mV). (d) Adsorption model of MnPc on
Bi(110).}\label{fig:fig1}
\end{figure}

Here, we show that the magnetic moment of a manganese phthalocyanine
(MnPc) molecule (see Fig.\,\ref{fig:fig1}(a)) on a Bi(110) surface is
reduced when  coordinated to a CO molecule. Using a combination of a low
temperature scanning tunneling microscopy (STM) and density functional
theory (DFT) we find, first, that the spin of the MnPc molecule is
reduced from $S=3/2$ to $S=1$ upon adsorption and, second, that CO further
reduces the spin of the MnPc from $S=1$ to $S=1/2$. The change in the
magnetic ground state upon CO attachment is detected by the
broadening of a characteristic zero-bias anomaly (ZBA). We interpret this broadening as a
transition from a Kondo regime, for the bare MnPc on Bi(110), to a regime
where charge fluctuations drive the CO-MnPc complex into a mix-valence
state. We further show that
the original magnetic state of MnPc is recovered after tip-induced
desorption of the CO molecule.

Our experiments have been performed in a custom-made scanning tunneling
microscope (STM) working in ultrahigh vacuum and at low temperature (5\,K). As a
semimetal, Bismuth  has a low density of states close to the Fermi level
\cite{Hofmann:2006}, favoring that  molecular adsorbates lie weakly affected by the surface  \cite{SchulzePRL12}. We chose the (110) surface because it additionally
presents dangling bonds that can eventually anchor the
molecular species to a fixed site, creating a full commensurate layer. We exposed an
atomically clean Bi(110) single crystal surface  at room temperature to a
flux of MnPc molecules (Sigma-Aldrich) thermally sublimed in vacuum from a crucible, and posteriorly cool it down for STM inspection.

On this surface MnPc molecules self-assemble in densely packed islands
(Fig.\ref{fig:fig1}(a)).  MnPc molecules appear as a clover-like
protrusion with a bright center corresponding to the metal ion. The
adsorption configuration could be established with precision  by resolving
simultaneously intramolecular structure of MnPc and atomic structure of
the underlying substrate (see e.g. Fig.\,\ref{fig:fig1}(c)). As shown
schematically in Fig.\,\ref{fig:fig1}(d), MnPc adsorbs on Bi(110) with its
Mn ion  located directly on top of a dangling bond of the surface. This
configuration suggests the formation of a local bond between MnPc and the
dangling bond of the bismuth atom.

The MnPc covered sample was subsequently annealed up to 130\,K and exposed
to CO gas partial pressure of $10^{-7}$\,mbar for 90\,s. After CO exposure
the structure of the molecular islands remains unchanged, but several
molecules ($\sim$10\%) exhibit now larger apparent height
(Fig.\,\ref{fig:fig1}(b)). We identify these new molecules as CO-ligated
MnPc. Figure\,\ref{fig:fig1}(c) shows a high resolution STM image of the
new species embedded in the MnPc island. Similarly to MnPc, the
CO-coordinated molecules exhibit fourfold symmetry, with a central
protrusion appearing 0.8\,\AA\ higher than in case of MnPc, indicating
that a single CO molecule bonds directly to the transition metal ion. We
find only singly coordinated CO-MnPc, contrary to previous studies reporting
doubly coordinated porphyrin molecules
\cite{Seufert:2010NAt,Seufert:2010JACS}.

\begin{figure}
\includegraphics[width=0.90\columnwidth]{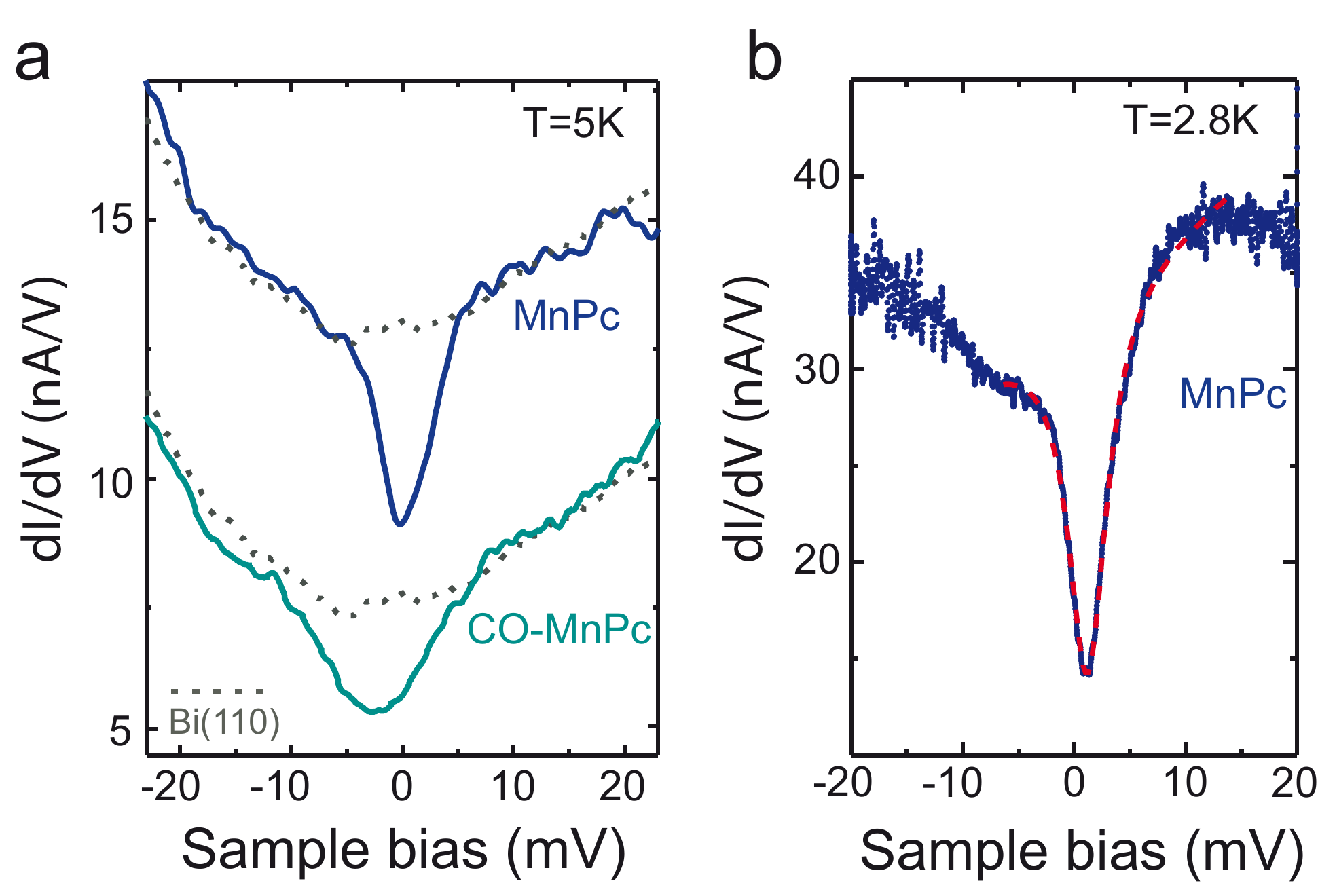}
\caption {(a) dI/dV spectra of
MnPc and CO-coordinated MnPc in the bias range close to E$_F$, showing the zero bias anomaly. The
dotted plots are the spectra measured on
bare Bi surface as a reference.
(b) Zero-bias feature of MnPc measured at reduced temperature T=2.8\,K.
The feature exhibits clear Fano line shape. Continuous line (red) corresponds to the fit by Fano equation.} \label{fig:fig2}
\end{figure}

The differential conductance ($dI/dV$) spectra measured close to E$_F$
reveals a pronounced anomaly at zero bias with a dip-like line shape on
MnPc (Fig.\,\ref{fig:fig2}(a)). We interpret it as a fingerprint of the
Kondo effect, as it has been observed before for MnPc and other
metal-phthalocyanines   on surfaces
\cite{Fu:2007,Gao:2007,Franke2011,Mugarza:2011}. The zero-bias anomaly
exhibits a clear Fano line shape (Fig.\,\ref{fig:fig2}(a) and (b)),
commonly observed for Kondo ground states of magnetic adsorbates
\cite{Ujsaghy:2000}. Other possible effect  which could also cause a zero-bias
anomaly, as e.g. inelastic spin flip excitation,  have been carefully
excluded by studying the temperature dependence of the line shape. A Kondo
temperature of T$_K^{MnPc}$=22($\pm6$)\,K can be approximated from a fit
to the anomaly's line shape using the Fano equation
\cite{Nagaoka:2002,footnote_Fano}.

\begin{figure}
\includegraphics[width=0.90\columnwidth]{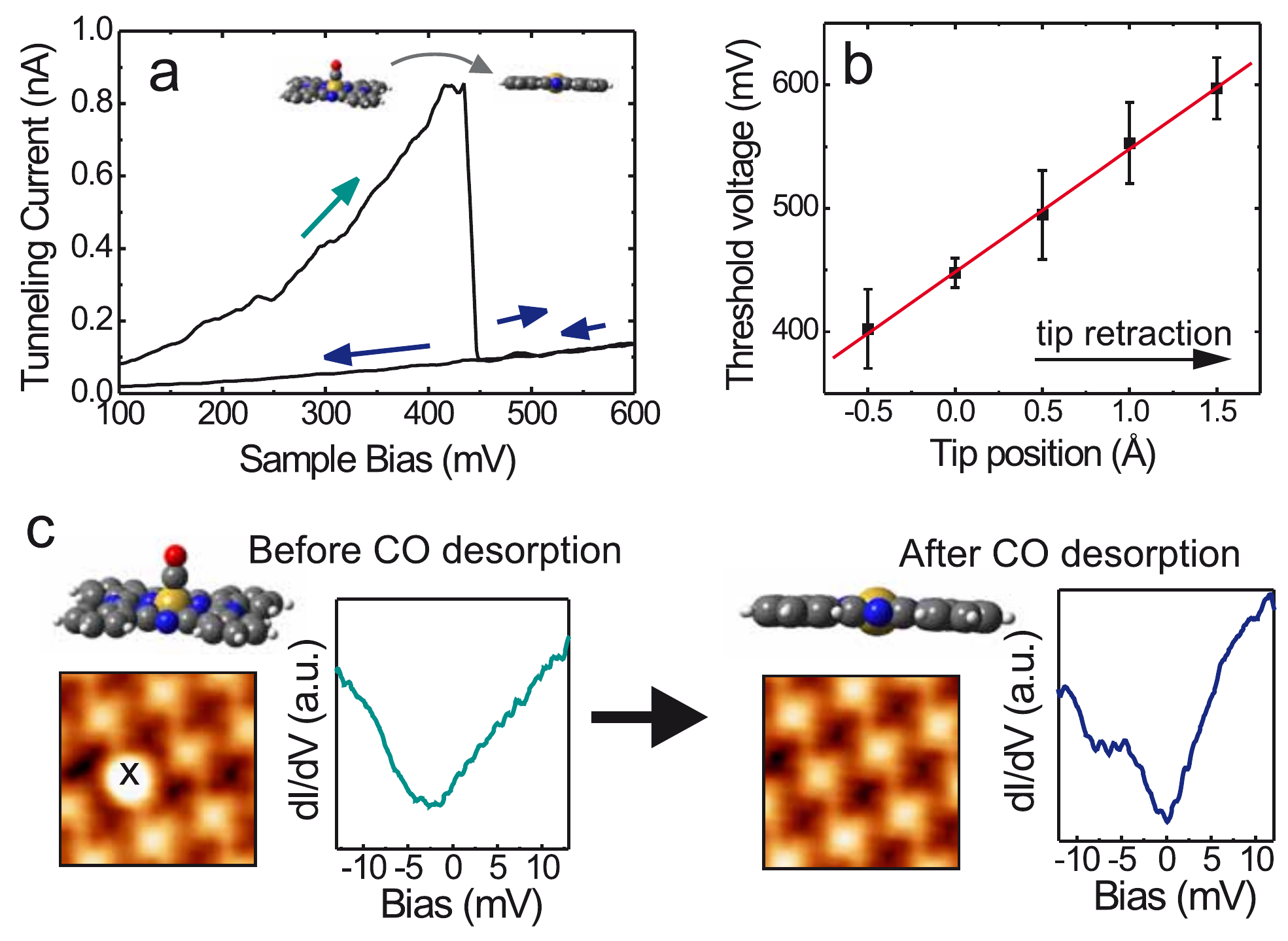}
\caption{ Desorption of CO molecules from MnPc islands. (a) I-V
characteristics recorded during desorption of CO. A sudden drop in current marks the threshold
voltage for CO detachment.  (b) Threshold
voltage plot versus tip-sample distance. $Zero$
position of the tip corresponds to the feedback parameters
$V_f$=100\,mV and $I_f$=100\,pA. Each point is an average of over
30 measurements. Linear fit of the data indicates that the process
is driven by electric field of 1\,V/nm. (c) STM image and dI/dV
spectrum of the molecule before and after controlled CO
desorption. } \label{fig:fig3}
\end{figure}

When similar spectra is measured on CO-MnPc molecules, the ZBA appears
much broader, with a slightly different lineshape
(Fig.\,\ref{fig:fig2}(a)). The broadening of the zero-bias resonance indicates
that the coordination to CO modifies the magnetic state of the molecule. This
modification is reversible and can be controlled by selective removal of
CO molecules using STM tip. In order to detach an individual CO from its
site, the STM tip was placed over CO-MnPc molecule, the feedback loop was
opened and the sample bias was ramped. The resulting current-voltage
characteristics is shown in Fig.\,\ref{fig:fig3}(a). A sudden drop of the
tunneling current at a certain threshold voltage indicates detachment of
the CO molecule from its coordination site (in  most cases to a new
site in a neighbor MnPc molecule). The resulting species recovers the
usual clove-like shape and a narrower zero-bias anomaly characteristic of
a bare MnPc molecule (Fig.\,\ref{fig:fig3}(c)). To establish the
underlying mechanism of CO detachment we note that the threshold voltage
increases linearly with the distance between the tip and the molecule, as
shown in Fig.\,\ref{fig:fig3}(b). This behavior indicates that the
desorption of CO is induced by the electric field at the tunnel junction.
A critical value of electric field  $\varepsilon \cong 1$\,V/nm can be
extracted from the slope of the line.

The origin of the zero-bias anomaly on the CO-MnPc molecules cannot be
deduced directly from our experiments. One could associate it to a
fingerprint of a Kondo ground state with a larger Kondo temperature
(T$_K^{CO-MnPc}$ about 50($\pm$10)\,K can be fitted). This is, however,
unprovable considering the $\pi$-acceptor character of the CO ligand. As a
strong field ligand, CO affects the magnetic state of MnPc in two ways:
(i) it increases the splitting of the $d$-orbitals, and thus favors the
low spin configuration in the complex \cite{Isvoranu:2011_CO}; (ii) it
withdraws the electron density from the bond to the ligand in the opposite
coordination site (here the Bi dangling bond) and hence reduces the
coupling of the $d$-orbitals to the substrate
\cite{Flechtner:2007,Gottfried:2009,Hieringer:2011}. The latter effect,
known in coordination chemistry as \textit{trans}-effect, should reduce
the Kondo temperature, contrary to our experimental observation.

\begin{figure}
\includegraphics[width=0.90\columnwidth]{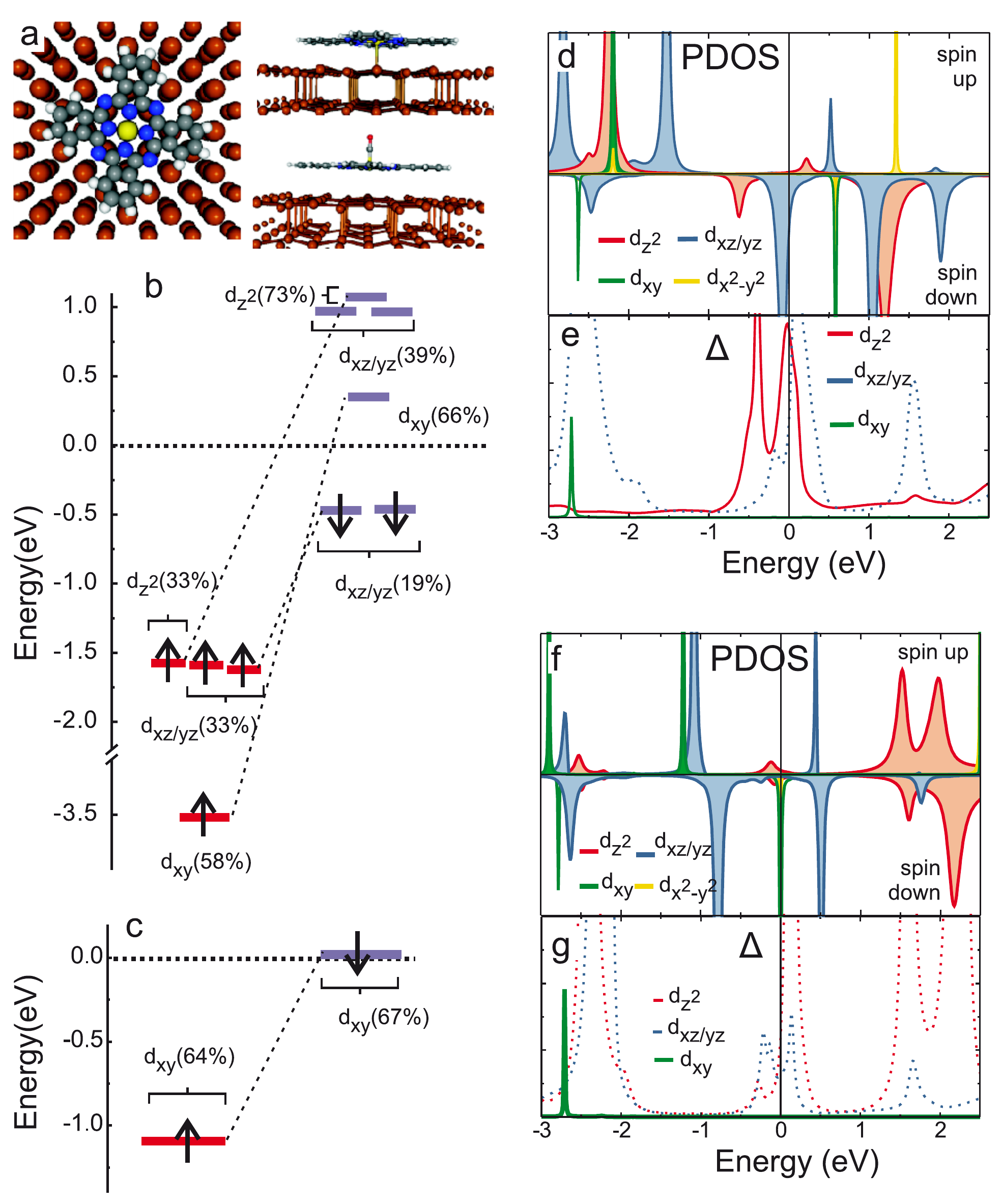}
\caption {(a) Model of computed system: top view (left) and side view (right) of
CO-MnPc (top) and MnPc (down). (b,c) Energy spectrum of adsorbed MnPc (b) and CO-MnPc (c),
showing the relevant spin states. Bars in left(right) show the up(down) eigenvalues.
The rate in parenthesis gives the atomic character of the molecular orbital. (d, f)
Local density of states (PDOS) and (e,g) hybridization function ($\Delta$) projected
on $d$-orbitals of adsorbed MnPc (d, e) and CO-MnPc (f, g). The dotted lines in (e)
and (g) refer to the orbitals which are not relevant for the magnetism of the molecule.}
 \label{fig:fig4}
\end{figure}

To rationalize the experimental results we performed density functional
theory (DFT) calculations of free and adsorbed MnPc and CO-MnPc molecules.
All calculations were performed using the Green's function formalism
implemented in our code ANT.G, which interfaces with Gaussian03/09
\cite{G03CODE,ANTCODE}. We used a general gradient approximation (GGA)
with the  Becke and Perdew-Becke-Ernzerhof exchange-correlation functional
\cite{GGA} in combination with a common double local basis set LANL2DZ
\cite{basis1,basis2}. Different views of the computed system are shown in
Fig.\,\ref{fig:fig4}(a). The Bi(110) surface was described using two
bilayers embedded on a tight-binding Bethe lattice. Following the
experimental results (Fig.\,\ref{fig:fig1}), MnPc and CO-MnPc molecules
were placed with the manganese atom directly on top of a Bi atom with an
optimized distance to the surface and rotated seven degrees with respect to the high symmetry axis of the Bi(110) surface. In this configuration, four carbon atoms of the phthalocyanine
cycle and the central Mn ion are in direct contact with Bi(110) dangling
bonds.

We simulate first the electronic  and spin
configuration of the free MnPc molecule, obtaining results in close agreement with previous
studies \cite{Wu09,ShenJCP10,Grobosch11}.
The molecule has a spin state $S=3/2$ due to three unpaired electrons
localized on the manganese orbitals $d_{xz}/d_{yz}$, $d_{z^2}$ and
$d_{xy}$. The $d_{xz}$/$d_{yz}$ orbitals, hosting three electrons, are
strongly hybridized with the phthalocyanine ring of the molecule, masking
the atomic character of its localized spin. The $d_{xy}$ orbital hosts the
most localized unpaired electron.

On the Bi(110) surface, MnPc keeps two
unpaired electrons in the $d$-orbitals.
The net spin density in
the $d_{xz}$ and $d_{yz}$ orbitals is reduced and results now from four
occupied molecular orbitals with different atomic contribution depending
on the spin orientation (Fig.\ref{fig:fig4}(b)). In addition, a spin
polarization arises in the ligands, opposite to that in the Mn core. Thus,
the total spin is reduced to $S=1$, with two identifiable and localized
unpaired electrons in the the $d_{z^2}$ and $d_{xy}$ orbitals
(Fig.\ref{fig:fig4}(b)).
We also find that a spin polarized density arises on Bi atoms directly
underneath the adsorbed molecule, 
 antiferromagnetically coupled to the spin on the Mn core and thus
supporting the emergence of Kondo screening.

The strength of a Kondo screening channel depends, to a first
approximation, on the hybridization with the substrate. For MnPc on
Bi(110), the spin-polarized local density (LDOS) of states projected on
the Mn-$d$ orbitals shows that each orbital hosting an unpair electron
interacts differently with the surface (Fig.\ref{fig:fig4}(d)). The
Mn-$d_{z^2}$ orbital hybridizes with a dangling bond of Bi and broadens
substantially. The $d_{xy}$ orbital  appears as narrow peaks in the
PDOS spectrum. This indicates that this state remains highly decoupled from the surface, probably to the absence of $z$-component. The result is that the two unpaired spins are screened via Kondo screening channels with very different energy
scale and, thus,  Kondo temperature \cite{Franke2011}. Such scheme
suggests an underscreened spin  if the experimental temperature lies
between the Kondo temperatures of each screening channels
\cite{Posazhennikova:2007,Korytar:2011}.

In the DFT framework, the orbital-substrate coupling is represented by a
hybridization function which allows for a qualitative analysis of the
feasible Kondo channels \cite{jacob2010,jacob2011}. The  hybridization
function is given by: $ \bfm{\Sigma_{d}}(\omega)=
(\omega+\mu)\bfm{I}-\bfm{{H}^{KS}_{d}}-\left(\bfm{G_{d}}(\omega)\right)^{-1}
$, where $ \omega $ and $\mu$ are the energy and the chemical potential
respectively, $\bfm{I}$ is the identity matrix, $\bfm{{H}^{KS}_{d}}$ is
the Kohn-Sham Hamiltonian and $\bfm{G_{d}} $ is the one electron green's
function projected onto d-subshell. Only the imaginary part of the
hybridization function is relevant in this work
$(\Delta=-Im\bfm{\Sigma_{d}}(\omega) / \pi)$. Specifically, the
prerequisite for the Kondo effect of the host orbital having a finite
coupling to the electron bath is equivalent to having non-zero value of
the hybridization function $\Delta$ at the Fermi level \cite{Hewson}. The
value of $\Delta$ calculated for different Mn $d$-orbitals of adsorbed
MnPc is shown in Fig.\,\ref{fig:fig4}(g). The $d_{z^2}$ orbital shows a
peak in the hybridization function at the Fermi level indicating that this
orbital is strongly screened  by the substrate. On the contrary, the
hybridization function  of the $d_{xy}$-orbital has a zero value in a
broad energy window in the vicinity of the Fermi level. These results
suggest that the Fano resonance in the spectra of the MnPc molecule is a
fingerprint of a Kondo channel opened to screen the $d_{z^2}$ orbital. The
unpaired spin of the $d_{xy}$-orbital remains decoupled from the electron
bath and this Kondo channel is closed.  On the basis of these simulations,
we conclude that the MnPc molecule retains part of its magnetic moment on
the surface, lying in an underscreened many-body ground state. This is a
consequence of the huge disparity in hybridization functions between spin
channels, making this system unique in this regard.

Next, we perform a similar analysis for the CO-MnPc molecular complex. The
CO molecule stays perpendicular to the MnPc molecular plane
(Fig.\,\ref{fig:fig4}(a)). It is anchored to the manganese ion via its
carbon atom through a synergic $\pi^{\ast}$-back-bonding, which strongly modifies the
electron population and the spin of the Mn $d$-orbitals in the following manner.
The $d_{z^2}$ and the
$d_{xz}/d_{yz}$ orbitals loose their magnetic moment: the former is
emptied due to the overlap with a nonbonding orbital of CO and the latter
is fully occupied due to the bond with the $p_x$ and the $p_y$ orbitals of
CO.
The spin-resolved LDOS
of the adsorbed CO-MnPc projected on the Mn $d$-orbitals
(Fig.\ref{fig:fig4}(f)) confirms that the spin-polarization of the
$d_{z^2}$ and $d_{xz}/d_{yz}$  orbitals is reduced.
Only the
$d_{xy}$-orbital appears to be relevant for the magnetism of the complex.
The total spin of the MnPc-CO complex is  reduced to
$S=1/2$, steaming from an unpaired electron  localized in a
molecular orbital with strong atomic $d_{xy}$-orbital character but a
finite contribution from the phthalocyanine cycle (see
Fig.\,\ref{fig:fig4}(c)).

It thus appears clear that the modification of the experimentally observed  Kondo resonance upon CO coordination is a direct consequence of the
redistribution of the $d$-electrons of the Mn atom. However, the ZBA  broadening cannot be understood as an increase in the Kondo screening because the responsible level
is very localized. As seen in
Fig.\,\ref{fig:fig4}(g), the  $d_{xy}$ hybridization function  is zero
in the vicinity of the Fermi level, what confirms
that also now  no Kondo effect can be mediated by this orbital.

This apparent inconsistency with the experimental observations can be
bypassed by noting that the PDOS on the  $d_{xy}$-orbital shows a highly
localized branch pinned at the Fermi level. This state thus belongs to the
lowest unoccupied orbital of the molecule (LUMO), and is the only
responsible for the chemical potential line-up between the molecule and
the substrate \cite{Ishii:1999}. In these circumstances, charge
fluctuations are expected to occur at the $d_{xy}$-orbital. When
electronic correlations are considered, a mixed-valence regime is likely
to emerge, as it is known for the $f$-electron compounds \cite{Hewson}. In
this situation, one expects a resonance to appear in the differential conductance spectra at zero bias, similar to that in the Kondo regime but characterized by a larger width
\cite{Costi:1994,Goldhaber-Gordon1998}. It is interesting to note that the small
hybridization with the surface prevents the finite spin in the
$d_{xy}$-orbital from quenching, in contrast with  the scenario reported
in related systems \cite{StepanowPRB11,MugarzaPRB12}. This allows for a many-body correlated state to emerge.

Finally, we comment on the possible role of the surface states in the
screening process on Bi(110). As a semimetal, bismuth is characterized by
a low density of bulk states at the Fermi level; its surfaces are however
strongly metallic due to the presence of the surface states. Therefore,
the Kondo effect on bismuth can be expected to involve surface states
rather than bulk states. However, strong spin-orbit coupling induces the
spin polarization of the surface states through the Rahsba effect
\cite{Hofmann:2006}; the concept of screening in such system is not
straightforward \cite{Zarea:2012}. The Kondo effect involving the chiral
spin-states has been recently intensively discussed in the context of
topological insulators \cite{Cha:2010,Feng:2010,Zitko:2010TI} and can be
understood assuming a complex structure of the Kondo cloud with a
nontrivial spatial and spin dependence \cite{Zitko:2010TI}.


In summary, we have demonstrated that the magnetic ground state of
individual metal-organic molecules can be reversibly modified by chemical
coordination to external molecular species acting as a ligands. Using a
combination of low-energy scanning tunneling spectra and ab initio
calculations, we find that the bonding of a CO molecule to the Mn$^{2+}$
center of MnPc molecules on Bi(110) leads to a substantial modification of its spin
state. Our results indicate that the chemically induced magnetic
transition corresponds to a change from a high-spin underscreened Kondo
state to a mixed-valence state, where charge fluctuations destroy the spin
1/2 state of the CO-MnPc complex. Our results thus emphasize the ultimate
limits of manipulating magnetic states of individual molecules by using a
combination of tunneling electrons and chemical control over the molecular
spin, which  could facilitate molecular scale patterning of magnetic
motifs in molecular thin films.

We thank K. J. Franke, B. W. Heinrich and D. Jacob for fruitful
discussions. This research was supported by Deutsche
Forschungsgemeinschaft (DFG-STR 1151/1-1 and SFB 658) and by MICINN under
grants FIS2010-21883 and CONSOLIDER CSD2007-0010. M. S. acknowledges
computational support from the CCC of the Universidad Aut\'onoma de
Madrid.


\begin{thebibliography}{99}
\bibitem{footnote_Fano} Thermal and instrumental broadening has been
    removed.
\end{thebibliography}

\end{document}